\documentclass[11pt,a4paper]{article}

\usepackage{amsmath}
\usepackage{amssymb}
\usepackage{graphicx}
\usepackage{dcolumn}
\usepackage{bm}
\usepackage{tabularx}
\usepackage[titles]{tocloft}
\usepackage{cite}
\usepackage{url}
\usepackage{xspace}
\usepackage{csquotes}
\usepackage{xcolor}

\title{Unifying dark matter, dark energy and inflation with a fuzzy dark fluid}
\author{A. Arbey, J.-F. Coupechoux}

\begin{document}

\begin{flushright}
CERN-TH-2020-117
\end{flushright}

\begin{center}
\Large\bf\boldmath
\vspace*{2.0cm}Unifying dark matter, dark energy and inflation\\[0.1cm] with a fuzzy dark fluid\\[1.cm]

\large One field to rule them all, and in the darkness bind them
\unboldmath
\end{center}

\vspace{0.5cm}
\begin{center}
A. Arbey$^{a,b,c,}$\footnote{Electronic address: \tt alexandre.arbey@ens-lyon.fr}, J.-F. Coupechoux$^{a,}$\footnote{Electronic address: \tt j-f.coupechoux@ipnl.in2p3.fr}\\[1.cm]

{\sl $^a$Univ Lyon, Univ Lyon 1, CNRS/IN2P3, Institut de Physique des 2 Infinis de Lyon UMR5822, F-69622 Villeurbanne, France\\[0.5cm]}

{\sl $^b$Institut Universitaire de France, 103 boulevard Saint-Michel, 75005 Paris, France\\[0.5cm]}

{\sl $^c$Theoretical Physics Department, CERN, CH-1211 Geneva 23, Switzerland\\[0.2cm]}

\end{center}
\vspace{1.5cm}
\begin{abstract}

Scalar fields appear in many cosmological models, in particular in order to provide explanations for dark energy and inflation, but also to emulate dark matter. In this paper, we show that it is possible for a scalar field to replace simultaneously dark matter, dark energy and inflation by assuming the existence of a non-minimal coupling to gravity, a Mexican hat potential, and a spontaneous symmetry breaking before inflation. After inflation, the scalar field behaves like a dark fluid, mimicking dark energy and dark matter, and has a dark matter behaviour similar to fuzzy dark matter.

\end{abstract}

\newpage

%%%%%%%%%%%%%%%%%%%%%%%%%%%%%%%%%%%

\section{Introduction}

According to the $\Lambda$CDM standard cosmological model, more than $95\%$ of the energy in the Universe is unknown, and the nature of dark energy and dark matter remains one of the most important and unresolved questions. The behaviour of dark matter and dark energy can be described using scalar fields. The only fundamental scalar field of the Standard Model of particle physics is the Higgs field, whose existence was experimentally confirmed by the discovery of the Higgs boson and measurements of its production and decay rates \cite{Aad:2012tfa,Chatrchyan:2012ufa}. The Higgs field is however not a viable candidate for dark energy or dark matter.

Dark energy was postulated to explain the recent acceleration of the expansion of the Universe. In the $\Lambda$CDM model the observations can be described using a simple cosmological constant $\Lambda$, equivalent to a constant dark energy density with a negative pressure. Such a constant density can also be obtained with a scalar field dominated by its potential. In this context quintessence models \cite{Zlatev:1998tr,Amendola:1999er}, which are based on scalar fields, aim at replacing the cosmological constant paradigm by a scenario with a dynamical field. The main  difference between the quintessence scenarios and the $\Lambda$CDM model is that the quintessence density is expected to evolve and could have played a role at earlier stages of the expansion of the Universe. The main issue is that the choice of the scalar field potential is still an open question.

Dark matter was postulated to explain extra gravitational effects which are observed at large astrophysical and cosmological scales, for example within the flat spiral galaxy rotation curves. Indeed the baryon matter density obtained via the luminosity function of spiral galaxies is inconsistent with the observed rotation curves. Such observations however would be compatible with the existence of an invisible matter, dark matter, which cannot originate from the Standard Model. Many models suggest that dark matter is made of weakly-interacting massive particles (WIMPs), direct or indirect dark matter detection experiments and collider searches have been so far unsuccessful in finding dark matter particles. Furthermore WIMP dark matter is expected to be cold, with low velocities, and galaxy simulations have revealed problems such as cuspy halos and missing satellites \cite{Irsic:2017yje}. Other models which have been suggested to explain the cosmological observations are the so-called fuzzy dark matter \cite{Hu:2000ke} and spintessence \cite{Boyle:2001du} models, in which dark matter is described by a scalar field with a very tiny mass of $m \sim 10^{-22}$ eV. At galactic scales the scalar field forms Bose-Einstein condensates which may constitute galaxy-sized dark matter halos. Such halos have been shown to reproduce the observed rotational curves \cite{Arbey:2003sj}.

Cosmological scalar fields are not used only to describe dark energy and dark matter. The chaotic inflation models are also based on scalar fields. These models have been introduced to solve two problems \cite{Guth:1980zm}: The first one is the horizon problem, which arises because the cosmological microwave background (CMB) is highly homogeneous even in apparently causally disconnected regions. The second one is the flatness problem, which requires a fine-tuning of the curvature parameter at the beginning of the Universe. These problems can be solved by assuming the existence of an exponentially-accelerated expansion in the early Universe, which may have been driven by scalar fields generating inflation.

The scalar fields are hence ubiquitous in cosmology, even if only one exists in the Standard Model of particle physics. One possibility to reduce the number of cosmological scalar fields is to unify dark energy and fuzzy dark matter using a single scalar field. This so-called dark fluid scalar field has already been considered in Refs.~\cite{Bilic:2001cg, Arbey:2005fn} and is a possible solution to both dark energy and dark matter questions. However, such a model is not unique and the choice of the scalar field potential \cite{Arbey:2006it, Arbey:2008gw} remains unclear. Nevertheless most of the potentials that have been considered in the literature can be approximated by the sum of a quadratic mass term leading to a matter behaviour and an approximately constant term giving a dark energy behaviour. 
A second possibility is to unify inflation and dark energy \cite{Linde:2002ws}. In this case the simplest model is based on a potential composed of a mass term and a constant term which leads to two stages of accelerated expansion: The first accelerated expansion occurs in the early Universe and the second one at the present epoch. 
A third possibility is to unify inflation and dark matter. In such a case the chaotic inflation scalar field does not decay completely during reheating, and the density surviving the incomplete decay can behave as dark matter \cite{Bastero-Gil:2015lga}. 

A step further would consist in a triple unification to explain simultaneously dark matter, dark energy and inflation with a unique scalar field. Such a scenario was for example studied in Refs.~\cite{Liddle:2008bm, Liddle:2006qz} where the standard chaotic inflation scalar field with a mass $m \sim 10^{-6}M_P$ survives after an incomplete decay. The key feature of this scenario is that the scalar field density remains negligible during the radiation-domination era and oscillate around a non zero-minimum after radiation domination. Another model studied in Ref.~\cite{DeSantiago:2011qb} relies on a non-canonical kinetic term.

In this article we present a more natural triple unification scenario based on a cosmological scalar field undergoing a symmetry breaking before inflation. The inflationary period will be similar to Starobinsky inflation \cite{Starobinsky:1980te}, and the resulting scalar perturbations in the new vacuum will behave as a dark fluid, unifying dark matter and dark energy.

In the following we will first review the observational constraints on dark fluid models, and discuss possible scalar field potentials and their advantages. In a second part, we will present a novel triple unification scenario, before concluding.

%%%%%%%%%%%%%%%%%%%%%%%%%%%%%%%%%%%

\section{Dark fluid model}
A scalar field which interacts only gravitationally with baryonic matter associated to a dominating quadratic mass term in the potential can behave as collisionless matter and be a dark matter candidate. One the other hand a scalar field with a mostly constant potential can explain the current acceleration of the expansion of the Universe. As discussed in Refs.~\cite{Arbey:2005fn,Bilic:2001cg,Peebles:2000yy} it is possible to explain the dark energy and dark matter behaviours with a unique dark fluid. We study the properties of such a scalar field in view of the observational constraints.

\subsection{Dark fluid model and observational constraints}

\subsubsection{Galactic scale: Fuzzy Dark Matter}
Most of the dark matter models involve particles which interact only very weakly with Standard Model particles. Such weakly-interacting massive particles (WIMPs) are still undiscovered at colliders and in dark matter detection experiments. An alternative possibility based on scalar fields, namely fuzzy dark matter \cite{Hu:2000ke}, has recently re-attracted some attention \cite{Irsic:2017yje}. At galactic scale such models can reproduce the flatness of galaxy rotation curves \cite{Arbey:2001qi}. The scalar field, associated to a quadratic potential with a mass $m$, can form a Bose-Einstein condensate in gravitational interactions with baryonic matter. The condensate constitutes a galactic halo with a typical size given by the Compton wavelength:
\begin{equation}
l_{\text{compton}}=\frac{h}{mc}\,.
\end{equation}   
For a typical halo of $10$ kpc, the mass $m$ is estimated to be $m\sim 10^{-23}$ eV \cite{Arbey:2006it}. Having such a small mass is an advantage for fuzzy dark matter since it does not suffer from the so-called cuspy halo and missing satellite problems \cite{Irsic:2017yje}. Furthermore a scalar field with in addition to a mass term a quartic term of coupling constant $\lambda$, can also condensate within a radius $L$, which can be of the order of the typical size of a cluster. The relation between $L$ and $\lambda$ is given by: 
\begin{equation}
\lambda = \frac{8\pi G m^4 L^2}{c^2}\,.
\end{equation}
For a typical cluster size $L \sim1$ Mpc, the value of the quartic term coupling is $\lambda \sim 10^{-89}$ \cite{Arbey:2006it}.

The fuzzy dark matter model has been studied in the context of galactic halos, in particular in Refs.~\cite{Hu:2000ke,Hui:2016ltb}. The evolution of an ultralight scalar field $\phi$ is given by the Klein-Gordon equation. In a static galaxy gravitational interaction can be globally described in the Newtonian limit with the Poisson equation. In addition, a nonrelativistic dispersion relation can be safely assumed for the wave equation. The wavefunction of the scalar field can be written under the form $\psi=A \exp (i\alpha)$, where $A$ is the probability amplitude and $\alpha$ the phase, so that $\phi=A \cos(mt - \alpha)$ \cite{Hu:2000ke}. The evolution equation reads:
\begin{equation}
i\left( \partial_t + \frac{3}{2}\frac{\dot{a}}{a} \right) \psi = \left( -\frac{1}{2m}\nabla^2 + m\Psi \right) \psi\label{Schroedinger}\,,
\end{equation}
and the Poisson equation:
\begin{equation}
\nabla^2 \Psi = 4\pi G \delta \rho_{\phi}\label{Poisson}\,,
\end{equation}
where $\Psi$ is the Newtonian potential, $\dot{a}/a$ is the Hubble parameter and $\delta \rho_{\phi} = m^2 \delta |\psi|^2/2$ is the energy density of the scalar field. Equation \eqref{Schroedinger} is the Schr{\"o}dinger equation for a self-gravitating particle in a Newtonian potential in an expanding Universe. In Ref.~\cite{Hu:2000ke} it is shown that ultralight particles with a mass $m\sim 10^{-22}$ eV lead to smooth and minimum-sized halos, and therefore provide a solution to the cuspy halo and galaxy satellite problems of standard cold dark matter scenarios. This value of the mass is also compatible with constraints from the Lyman-$\alpha$ forest data and hydrodynamic simulations \cite{Irsic:2017yje}. 

The Schr{\"o}dinger-Poisson equation system (\ref{Schroedinger},\ref{Poisson}) also describes a solitonic behaviour during a head-on collision between two galaxies with Bose-Einstein condensate halos \cite{Bernal:2006ci}, that is compatible with the observations of the Bullet Cluster \cite{Clowe:2006eq}.

Galaxy rotation curves in agreement with the observations can also be obtained with an ultralight complex scalar field, when a stationary and regular configuration is assumed, with the harmonic ansatz:
\begin{equation}
\phi(r,t) = \phi_0(r)\exp (i\omega t)\,,
\end{equation} 
similar to the one of boson stars. A comprehensive study in general relativity has been presented in \cite{Arbey:2001qi}, finding a best fit to the galaxy rotation curves with a mass in the range $10^{-24}-10^{-23}$ eV. This result is similar to the one obtained in the Newtonian approximation.  

\subsubsection{Cosmological behaviour}
Let us now consider the cosmological behaviour of the dark fluid scalar field. We assume a homogeneous Universe filled only with radiation, baryonic matter and dark fluid scalar field. Using the Friedmann-Lemaître-Robertson-Walker (FLRW) metric, the Einstein and Klein-Gordon equations become:
\begin{equation}
\begin{aligned}
& H^2 = \frac{8\pi G}{3} \left( \rho_{\phi} + \rho_{r} + \rho_{b} \right)\,,\\
& 2\dot{H} + 3H^2 = -8\pi G \left( P_{\phi} + P_{r} + P_{b} \right)\,,\\
& \ddot{\phi} + 3H\dot{\phi} + \frac{dV}{d\phi} = 0\,,
\end{aligned}
\label{KGE}
\end{equation}
where the energy density and pressure of the scalar field are given by:
\begin{equation}
\begin{aligned}
&\rho_{\phi} = \frac{1}{2}\left( \frac{d\phi}{dt} \right)^2 + V(\phi)\,, \\
&P_{\phi} = \frac{1}{2}\left( \frac{d\phi}{dt} \right)^2 - V(\phi)\,.
\end{aligned}
\end{equation}

During the period when radiation is dominating the expansion, we can derive constraints from primordial nucleosynthesis (BBN). At this epoch the evolution of the scalar field density is governed by its kinetic term and BBN constraints exclude large scalar field densities such as \cite{Arbey:2019cpf}
\begin{equation}
\rho_{\phi}(1 {\rm\;MeV}) \geq 1.40 \rho_{\gamma}(1 {\rm\;MeV})\,.
\end{equation} 

The CMB and large scale structure observations provide additional constraints. Considering a scalar field with anharmonic corrections and a potential with a quadratic term with a mass $m$ and a quartic term with a small coupling constant $\lambda$, the main constraints provided by the Planck and WiggleZ data  are \cite{Cembranos:2018ulm}:
\begin{equation}
\log_{10}(\lambda) < -91.86 + 4\log_{10}\left( \frac{m}{10^{-22} \text{ eV}} \right)
\label{constraint_lambda}
\end{equation}  
for masses heavier than $10^{-24}$ eV.

The acceleration of the expansion of the recent Universe can be in agreement with a simple cosmological constant. On the other hand quintessence models describe dark energy with a scalar field $\phi$ which has a constant density today and was negligible during the matter and radiation domination periods. The Planck Collaboration provides constraints on the equation of state $w_{\phi} = P_\phi/\rho_\phi$ \cite{Aghanim:2018eyx}. For the two parametrizations proposed in~\cite{Linder:2005ne}, the constraints are the following:
\begin{itemize}
\item if $w_{\phi}$ is constant,
\begin{equation}
w_{\phi} = -1.028 \pm 0.032,
\end{equation}  
\item if $w_{\phi}(a) = w_0 + (1-a) w_a$, where $a$ is the expansion factor, and $w_0$ and $w_a$ two constants:
\begin{equation}
\begin{aligned}
&w_0 = -0.961 \pm 0.077\,,\\
&w_a =-0.28^{+0.31}_{-0.21}\,.
\end{aligned}
\end{equation}   
\end{itemize}
To have a dark energy behaviour, the dark fluid scalar field therefore needs to have in the recent Universe a rather constant density, corresponding to a nearly constant value of $\phi$.

\subsection{Polynomial dark fluid models}

\subsubsection{Average evolution}
We consider the cosmological evolution of a rapidly oscillating scalar field with a frequency $f_{\rm eff}$. The oscillations need to be faster than the Universe evolution, which is characterized by the conformal Hubble time $\mathcal{H}^{-1}$. For example for the power-law potential $U(\phi)=\lambda |\phi|^n/n$, the average equation of state is given by
\begin{equation}
w_\phi = \frac{\langle P_\phi \rangle}{\langle \rho_\phi \rangle} = \frac{n-2}{n+2} \,,
\end{equation} 
where $\langle...\rangle$ denotes the average value over a time $T$ such that $\mathcal{H}^{-1}\gg T\gg f_{\rm eff}^{-1}$. When the potential governs the energy density evolution, the conservation of the stress energy tensor gives
\begin{equation}
    \langle \rho_{\phi} \rangle = \rho_{\phi,0} \, \left(\frac{a}{a_0}\right)^{-3(1+w_\phi)} =\rho_{\phi,0} \, \left(\frac{a}{a_0}\right)^{-6n/(n+2)}\,.
\label{rho}
\end{equation}

\subsubsection{Simple dark fluid potential}
\label{sec:simple}

The simplest potential for a dark fluid model is the sum of the fuzzy dark matter potential and a cosmological constant:
\begin{equation}
V(\phi)=V_0+\frac{1}{2}m^2\phi^2\,.
\label{simple_pot}
\end{equation} 
The constant $V_0$ can lead to a dark energy behaviour provided
\begin{equation}
V_0 = \frac{\Lambda c^4}{8\pi G} \approx 2.5\times 10^{-11} \text{ eV}^4\,,
\end{equation}
where $\Lambda$ is the cosmological constant. The mass term behaves as a cold and fuzzy dark matter if $m$ is in the range:
\begin{equation}
m \approx 10^{-22}-10^{-21} \text{ eV}\,.
\end{equation} 

\begin{figure}[!th]
\centering
\includegraphics[width=15cm]{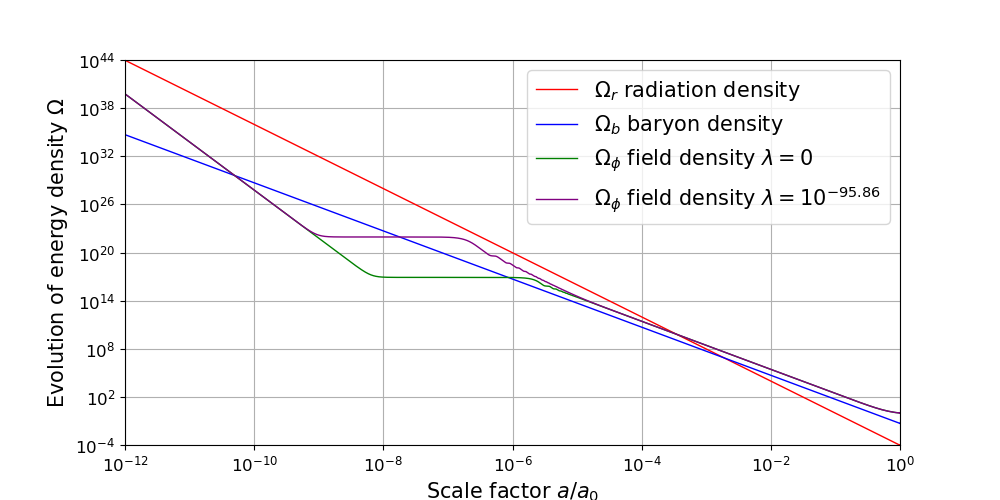}
\caption{Evolution of dark fluid scalar field density fraction as a function of the scale factor $a$ in the case of a potential with constant term, quadratic term  with mass $m=10^{-22}$ eV, and quartic term with coupling constant $\lambda$. Baryon and radiation densities follow those of the $\Lambda$CDM model.\label{fig:simple_dark_fluid}}
\end{figure}

Figure~\ref{fig:simple_dark_fluid} shows the evolution of the scalar field energy density as a function of the scale factor $a$ in a homogeneous Universe described by the system of equations~\eqref{KGE}. More precisely the Universe is considered flat and composed of:
\begin{itemize}
\item radiation which evolves according to $a^{-4}$,
\item baryonic matter which evolves according to $a^{-3}$,
\item dark fluid which evolves like matter as $a^{-3}$ when the mass term governs the scalar field evolution and like dark energy when the constant term dominates.
\end{itemize}
Within this setup, the dark fluid can replace simultaneously dark matter and dark energy.

\subsubsection{Quadratic potential with anharmonic corrections}
\label{sec:quartic}
Considering now one extra order in the polynomial expansion the potential reads
\begin{equation}
V(\phi) = V_0 + \frac{1}{2} m^2 \phi^2 + \frac{1}{4} \lambda \phi^4\,.  
\end{equation}
In this case the scalar field evolves like:
\begin{itemize}
\item radiation when $m^2 \ll \lambda \phi^2$ and $V_0 \ll \lambda \phi^4$\,,
\item cold dark matter when $\lambda \phi^2 \ll m^2$ and  $V_0 \ll m^2 \phi^2$\,,
\item cosmological constant when $\lambda \phi^4 \ll V_0$ and $m^2 \phi^2 \ll V_0$\,.
\end{itemize}
The constant term  $V_0$ of the potential is in general arbitrary and we take it equal to the cosmological constant.

The cosmological evolution of the scale factor in a Universe made of baryonic matter, radiation and scalar field is given by:
\begin{equation}
\begin{aligned}
\frac{\dot{a}}{a} &= H_0 \sqrt{\frac{\Omega_r^0}{a^4} + \frac{\Omega_b^0}{a^3} + \frac{1}{\rho_{cr}^0}\left(\frac{\dot{\phi}^2}{2} + V\right)}\,,\\
\ddot{\phi} &= -3 \frac{\dot{a}}{a} \dot{\phi} - \frac{d V}{ d\phi}\,,
\end{aligned}
\end{equation}
with $\rho_{cr}^0=3H_0^2/(8\pi G)$ the critical density and $\Omega_X=\rho^0_X/\rho_{cr}^0$. In the early Universe the scalar field evolution is dominated by the kinetic term $\dot{\phi}^2/2$ and its density evolves as $a^{-6}$. After this period the potential can start dominating, resulting in a constant density. Then an equilibrium between the kinetic and mass terms is reached, leading to a matter behaviour. Essentially the value of the constant density plateau depends on on the mass $m$, that we fix to $10^{-22}$ eV, the initial value of $\phi$ and the quartic term . The dependence on $\lambda$ is shown in Fig.~\ref{fig:simple_dark_fluid} for a quadratic potential with anharmonic corrections, where the initial conditions have been chosen in order to have the same scalar field in the recent Universe. At later times the scalar field oscillates quickly and reaches the average solution. Usually the fuzzy dark matter potential is just a mass term with in some cases an anharmonic correction term. Yet the potential can be much more complicated and the usual fuzzy dark matter potential can correspond to the first orders of the Taylor expansion of a more general potential.

\subsection{General potential for the dark fluid model}

A possibility to improve the dark fluid model is to replace the constant part of the potential by a dynamical term. Practically it has to be negligible during the matter domination era in order not to affect the dark matter behaviour. One can for example add to the quadratic term a quintessence potential \cite{Tsujikawa:2013fta,Caldwell:2005tm}, replace the quadratic term \cite{Arbey:2006it}, consider quantum corrections \cite{Arbey:2007vu}, ... We will see in the following which potentials can explain simultaneously dark energy and dark matter.  

\subsubsection{Extended dark matter term}

Let us consider the simple case of an exponential potential. By construction: 
\begin{equation}
\begin{aligned}
V(\phi) &= V_0 \exp \left(\frac{m^2}{2V_0}\phi^2\right)\\
        & \underset{\phi\to0}{\simeq} V_0  + \frac{1}{2} m^2 \phi^2 + \frac{m^4}{4V_0} \phi^4\,,
\end{aligned}
\end{equation}
where the second equality is the Taylor expansion, which is valid for small $\phi$ values. $V_0$ is fixed to the cosmological constant value $V_0=2.5\times 10^{-11}$ eV, and $m\sim 10^{-22}$ eV as explained in the previous section, so that the effective quartic term coupling is
\begin{equation}
\lambda = \frac{m^4}{V_0} \simeq 10^{-78} \,.
\end{equation}
Unfortunately this value is too large and is not compatible with the CMB constraints given in Eq.~\eqref{constraint_lambda}.

A second possibility is the following potential:
\begin{equation}
\begin{aligned}
V(\phi) &= V_0 + \frac{m^4}{\lambda} \left[\exp \left( \frac{\lambda \phi^2}{2m^2} \right) -1 \right]\\
        & \underset{\phi\to0}{\simeq} V_0 + \frac{1}{2} m^2 \phi^2 + \frac{1}{4} \lambda \phi^4\,,
\end{aligned}
\end{equation}
which has the same behaviour as the polynomial dark fluid model presented in the previous section.

Another possibility is a potential with a hyperbolic sine:
\begin{equation}
\begin{aligned}
V(\phi) &= V_0 + \frac{m^4}{\beta} \sinh \left( \frac{\beta \phi^2}{2m^2}\right)\\
        & \underset{\phi\to0}{\simeq} V_0 + \frac{1}{2} m^2 \phi^2 + \frac{1}{4!} \frac{\beta^2}{m^2} \phi^6\,.
\end{aligned}
\end{equation}
When $\phi$ is small, there is no quartic term and the $\phi^6$ term will have a negligible effect, so that this potential reduces to a constant and a quadratic term.

These models have similar behaviours, which correspond to the one of a polynomial potential. In practice the main difference between them is the position of the constant density plateau, as explained in~\cite{Arbey:2019cpf}. After the plateau, the scalar field behaves as dark matter and/or as dark energy, and these models are indistinguishable.

\subsubsection{Extended dark energy term}

We consider the case in which a quintessence term is added to the dark fluid potential. There exist two main classes of quintessence models \cite{Caldwell:2005tm}: freezing models in which the field follows a tracker potential down to its minimum, and thawing models in which the field continues evolving after reaching the dark energy behaviour.

To have a dark fluid model it is necessary for the quintessence term not to modify the dark matter behaviour. For example the tracking freezing potential \cite{Linder:2006uf}
\begin{equation}
V(\phi) = M^4 \left( \frac{M_P}{\phi} \right)^p 
\end{equation}
is incompatible with the mass term. Indeed if the scalar field oscillates with a density evolving as $a^{-3}$, the field $\phi$ will sooner or later reach a negligible value and the tracking freezing potential diverges. 

We now consider the expansion of an exponential potential:
\begin{equation}
\begin{aligned}
V(\phi) &= \frac{1}{2} m^2\phi^2 + \alpha\exp \left(-\frac{\beta}{2}\phi^2\right)\\
        & \underset{\phi\to0}{\simeq} \alpha + \frac{1}{2}(m^2-\alpha \beta)\phi^2 + \frac{\alpha \beta^2}{4} \phi^4\,,
\end{aligned}
\end{equation}
where $\alpha$ is determined by the dark energy density, and $m$ and $\beta$ by the galactic and cluster scales, respectively. Therefore the exponential term can explain simultaneously the acceleration of the expansion, galaxy rotation curves and galaxy cluster scale. Numerically, the modification of the mass term by the $\alpha \beta$ term is tiny.
 
\begin{figure}[!t]
\centering
\includegraphics[width=15cm]{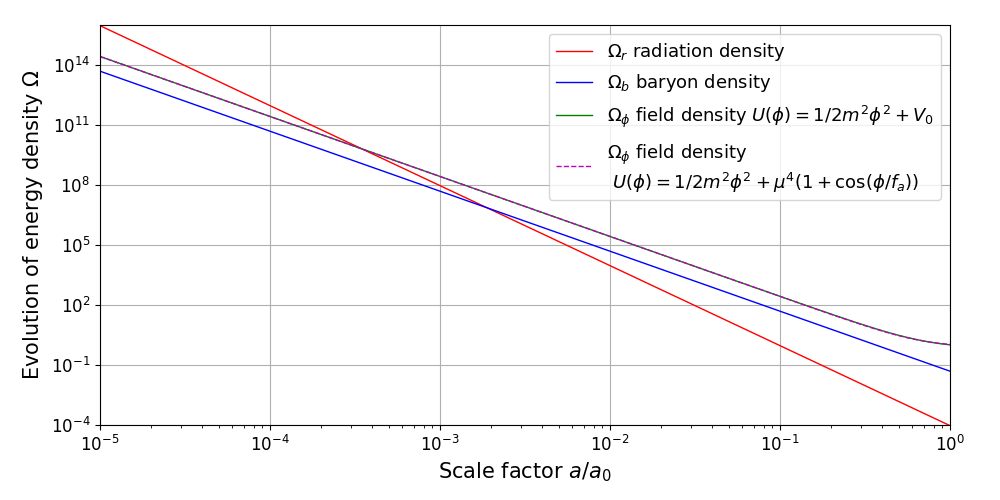}
\caption{Evolution of the dark fluid density for the simple dark fluid potential (solid green line) and for the pseudo-Nambu-Goldstone potential (dashed purple line) with mass $m=10^{-22}$ eV. Both curves are superimposed with negligible differences. The evolutions of the baryon density and radiation density are the same as in the $\Lambda$CDM model, and are shown in red and blue, respectively.\label{fig:extended_dark_energy}}
\end{figure}

Another possibility is the addition of a pseudo-Nambu-Goldstone potential \cite{Frieman:1995pm}:
\begin{equation}
V(\phi) = \frac{1}{2} m^2\phi^2 + \mu^4 \left( 1 + \cos (\phi/f_a) \right)\,,
\end{equation}
where $m$, $\mu$ and $f_a$ are constant parameters. In this thawing model the dark energy behaviour which presently dominates the evolution of the Universe will end in the future and a new period of dark matter domination will happen in the far future. In Figure~\ref{fig:extended_dark_energy}, the density evolution of the scalar field governed with the pseudo-Nambu-Goldstone potential with mass term is the same as with the simple dark fluid potential defined in Eq. \eqref{simple_pot}.

The dark fluid model can also be generalized to a complex scalar field with a $U(1)$ symmetric potential. The dark matter behaviour is based on the spintessence model \cite{Boyle:2001du}. The symmetry implies the conservation of a charge per comoving volume: $Q=R^2\dot{\theta}a^3$ where $R$ and $\theta$ are the amplitude and phase of the complex scalar field such as $\phi=R\exp(i\theta)$. However, similarly to all the dark fluid models that we presented, the choice of the potential remains arbitrary. The crucial point for a successful model seems to be that the dark fluid needs to have a fuzzy dark matter behaviour, or in other words its potential should contain a mass term with $m\sim 10^{-22}$~eV. 

%%%%%%%%%%%%%%%%%%%%%%%%%%%%%%%%%%%

\section{Triple unification}

In this section we build a model in which inflation can be simultaneously explained within a dark fluid model, in which the scalar field is non-minimally coupled to gravity. We restrict our study to cases where the dark fluid potential is a polynomial of order 4.

\subsection{Generalities}

\subsubsection{Chaotic inflation}
Inflationary models have been introduced to explain the flatness and horizon problems. Successful models are based on scalar fields with a slow-roll evolution during a sufficiently long time in order to provide a very large expansion rate in the early Universe, such that $N=\log(a_{end}/a_{beg}) \gtrsim 50$ \cite{Akrami:2018odb}, where $N$ is the number of e-folds. The so-called chaotic inflation model \cite{Linde:1983gd}, in which a scalar field with quadratic potential is in slow-roll during inflation, can fulfil this constraint. Also in Ref.~\cite{Linde:2002ws}, it is shown that chaotic inflation can be obtained together with dark energy using a scalar field associated to the potential
\begin{equation}
V(\phi) = \frac{m^2}{2} \phi^2 + V_0\,. 
\end{equation}
The main problem to unify chaotic inflation and dark energy is that the observation of the CMB anisotropies imposes $m \simeq 3 \times 10^{-6} M_P$ \cite{Aghanim:2018eyx}, which is not in agreement with the mass needed for a dark fluid model. Using instead a quartic potential which becomes dominant during the inflation:
\begin{equation}
V(\phi) \simeq \frac{\lambda}{4!} \phi^4\,,
\end{equation}
$\lambda$ has to be of the order of $10^{-14}$ to be compatible with CMB data \cite{Aghanim:2018eyx}, which is incompatible with the dark fluid setup. Therefore the dark fluid potentials derived in the previous section cannot lead to a chaotic inflation in agreement with the observational data.

\subsubsection{Non-minimal coupling $\phi^2 R$}
A first solution to unify inflation and dark fluid model would be to consider a non-minimal coupling between the dark fluid scalar field and the scalar curvature. Such couplings have been studied in the context of the Higgs-inflation scenario \cite{Bezrukov:2007ep}. Let us consider the following action:
\begin{equation}
\mathcal{S}=\int d^4x \sqrt{-g}\left[\frac{1}{2\kappa^2}\left(1+\frac{\alpha^2}{M_P^2}\phi^2\right)R-\frac{1}{2}g^{\mu \nu}\partial_{\mu}\phi\partial_{\nu}\phi-V(\phi) \right]\,,
\label{S2_1}
\end{equation}
with $\kappa^2 = M_P^{-2}$ and
\begin{equation}
V(\phi) = V_0 + \frac{m^2}{2}\phi^2 + \frac{\lambda}{4}\phi^4\,. 
\label{V_conf}
\end{equation}
The $\alpha^2$ coupling is chosen to be positive in order to ensure that the coupling to gravity always remains positive. The parameters $V_0$, $m$ and $\lambda$ are fixed by the dark fluid model requirements, and the only free parameter is therefore $\alpha$. This action is assumed to be written in the Jordan frame in which the scalar field is non-minimally coupled to the Ricci scalar $R$. To confront this model with the CMB data, it is necessary to rewrite the action in the Einstein frame by making a conformal transformation. In the Einstein frame, where the quantities are represented by a tilde, the metric is:
\begin{equation}
\tilde{g}_{\mu \nu} = \Omega^2 g_{\mu \nu}\,,
\end{equation}
where $\Omega^2$ is the conformal factor such that
\begin{equation}
\Omega^2 = 1 + \alpha^2 \frac{\phi^2}{M_P^2}\,.
\end{equation}
Following Ref.~\cite{Kaiser:1994vs}, one defines the effective scalar field $\psi$ and potential $U$ such that
\begin{equation}
\begin{aligned}
& \frac{d\psi}{d\phi} =  \sqrt{\frac{\Omega^2 + 6\alpha^4 \phi^2/M_P^2}{\Omega^4}}\,,\\
& U(\psi) = \Omega^{-4} V(\phi)\,,
\end{aligned}
\end{equation}
and the action \eqref{S2_1} takes the form of the usual Einstein-Hilbert action:
\begin{equation}
\mathcal{S}=\int d^4x \sqrt{-\tilde{g}}\left[\frac{1}{2\kappa^2}\tilde{R}-\frac{1}{2}\tilde{g}^{\mu \nu}\partial_{\mu}\psi\partial_{\nu}\psi-U(\psi) \right]\,.
\end{equation}
The observational constraints on the spectral index $n_s$ implies $\alpha^2>4\times 10^{-3}$ \cite{Akrami:2018odb}.

On the other hand the parameter $\alpha$ can be constrained by the observational power spectrum which is related to the potential by \cite{Lyth:1998xn}:
\begin{equation}
\delta_H^2 = \frac{4}{25} \mathcal{P}_{\mathcal{R}} = \frac{1}{150 \pi^2 M_P^4} \frac{U}{\epsilon_v}\,,
\end{equation}
where
\begin{equation}
\epsilon_v = \frac{M_P^2}{2}\left(\frac{U'(\phi)}{U(\phi)}\right)^2 = \frac{M_P^2}{\phi^2} \frac{8}{1+(1+6\alpha^2)\alpha^2\phi^2/M_P^2}
\end{equation} 
is the slow-roll parameter \cite{Kaiser:1994vs}. The power spectrum has to be calculated at the time of the end of inflation $t_{end}$, which is related to the number of e-folds $N$:
\begin{equation}
N = \int_{\phi_{beg}}^{\phi_{end}} \frac{\phi}{M_P} \frac{1+(1+6\alpha^2)\alpha^2\phi^2/M_P^2}{4\left(1+\alpha^2\phi^2/M_P^2\right)} d\phi\,,
\end{equation}
where $\phi_{end}$ corresponds to $\epsilon_v=1$. Unfortunately, the system:
\begin{equation}
\begin{aligned}
& \frac{M_P^2}{\phi_{end}^2} \frac{8}{1+(1+6\alpha^2)\alpha^2\phi_{end}^2/M_P^2} = 1\,, \\
& 8N = (1+6\alpha^2)\left(\frac{\phi_{beg}^2}{M_P^2}-\frac{\phi_{end}^2}{M_P^2}\right) + 6\ln \left( \frac{1+\alpha^2\phi_{beg}^2/M_P^2}{1+\alpha^2\phi_{end}^2/M_P^2}\right)\,, \\
& \mathcal{P}^*_{\mathcal{R}} = \frac{\lambda}{192\pi^2} \frac{\phi_{beg}^6}{M_P^6} \left(1+(1+6\alpha^2)\alpha^2\frac{\phi_{beg}^2}{M_P^2}\right)\,,
\end{aligned} 
\end{equation} 
has no valid solution for $\lambda \sim 10^{-100}$. For example assuming $\alpha^2 \ge 1$ the first equation leads to $\alpha^2\phi_{end}^2 \sim 1.2 M_P^2$, the second one gives $\alpha^2\phi_{beg}^2 \sim 75M_P^2$ for $N\simeq 55$, and the third one imposes $\alpha^2 \sim 10^{-42}$, which contradicts the other requirements. Therefore, it is not possible to unify inflation and dark fluid with a coupling $\phi^2R$. This result is different from the ones obtained in the context of Higgs-inflation model, in which the quartic coupling $\lambda$ is much larger. In our case the parameter $\lambda$ is too small to reproduce the amplitude of the anisotropies.

\subsubsection{Non-minimal coupling $\phi^2R^2$}

A third possibility is to add to the action a non-minimal gravitational coupling to the scalar curvature squared: $\phi^2 R^2$. Let us consider the following action:   
\begin{equation}
\mathcal{S}=\int d^4x \sqrt{-g}\left[\frac{1}{2\kappa^2}\left(R+\alpha \phi^2 R^2 \right)-\frac{1}{2}g^{\mu \nu}\partial_{\mu}\phi\partial_{\nu}\phi-V(\phi) \right]\,,
\label{S1}
\end{equation}
where $V$ is the dark fluid potential. In this model inflation has to occur when the $\alpha\phi^2 R^2$ term drives the evolution. Our goal is to retrieve the Starobinsky inflation model \cite{Starobinsky:1980te}. Before discussing the properties of the action \eqref{S1}, we briefly review the Starobinsky model, which is a specific case of $f(R)$ theories \cite{Sotiriou:2008rp}, in which the geometrical action reads
\begin{equation}
\mathcal{S}=\int d^4x \sqrt{-g} \frac{1}{2\kappa^2} f(R)\,,
\end{equation}
with
\begin{equation}
f(R) = R + \frac{R^2}{6 M^2}\,,
\end{equation}
where $M$ is a mass parameter. In the FLRW metric the Einstein equations lead to:
\begin{equation}
\begin{aligned}
& \ddot{H} - \frac{\dot{H}^2}{2H} + \frac{1}{2}M^2 H + 3H\dot{H} = 0\,,\\
& \ddot{R} + 3H\dot{R} + M^2R = 0\,.
\end{aligned}
\label{eq:R2}
\end{equation}
Assuming that inflation occurs when $R \gg M$ and $\dot{H} \gg H^2$ one can show that \cite{DeFelice:2010aj,Mijic:1986iv}:
\begin{equation}
\begin{aligned}
& H(t) = H_i - \frac{M^2}{6} (t-t_i) \,,\\
& R(t) = 12 H^2 - M^2 \,,\\
& a(t) = a_i \exp \left(H_i(t-t_i) - \frac{M^2}{12}(t-t_i)^2 \right) \,,
\end{aligned}
\label{Sol_R2inf}
\end{equation}
so that the Universe experiences an exponential inflationary expansion.

In order to find a similar behaviour with the action \eqref{S1}, $\alpha \phi^2$ needs to be constant and equal to $1/6M^2$. However the scalar field is dynamical and follows the Klein-Gordon equation, which reads in the FLRW metric:
\begin{equation}
\ddot{\phi} + 3H\dot{\phi} +V_{,\phi} - \frac{\alpha \phi R^2}{\kappa^2} = 0 \,,
\end{equation} 
and for which the only solution with constant $\phi$ corresponds to $\phi = 0$ with the minimal dark fluid potential.

To obtain a non-zero and constant $\phi$, a possibility is to consider a potential with a non-standard minimum which does not sit at $\phi=0$. We will study in the following the potential:     
\begin{equation}
V(\phi) = V_0 + \frac{m^2}{8v^2}\left(\phi^2-v^2\right)^2\,,
\label{V_sym}
\end{equation}
which has two minima corresponding to $\phi^2=v^2$. With this potential $\phi=\pm v$ is a constant solution to the Klein-Gordon equation, and the action \eqref{S1} can lead to inflation, dark matter and dark energy behaviours.

\subsection{$R^2$ inflation}

\subsubsection{$Z_2$ spontaneous symmetry breaking}

The action \eqref{S1} with the potential defined in Eq.~\eqref{V_sym} is invariant under a $Z_2$ symmetry ($\phi(x) \rightarrow -\phi(x)$). At $\phi = 0$ the potential has a local maximum and the theory is unstable around this value. The two minima correspond to $\phi=\pm v$. When the scalar field goes to one of these minima, the $Z_2$ symmetry is spontaneously broken. The scalar field $\phi$ can oscillate around one of the minima so that:
\begin{equation}
\phi = \xi \pm v\,,
\end{equation}
where $v$ is a VEV and $\xi$ is the variation of scalar field around the minimum. The action \eqref{S1} thus becomes:
\begin{equation}
\mathcal{S}=\int d^4x \sqrt{-g}\left[\frac{1}{2\kappa^2}\left(R+\alpha v^2 \left(1 \pm \frac{2}{v} \xi + \frac{1}{v^2} \xi^2\right) R^2 \right)-\frac{1}{2}g^{\mu \nu}\partial_{\mu}\xi\partial_{\nu}\xi- V(\xi) \right] \,,
\label{S2}
\end{equation}
with:
\begin{equation}
V(\xi) = V_0 + \frac{m^2}{2}\xi^2 \pm \frac{m^2}{2v}\xi^3 + \frac{m^2}{8v^2}\xi^4 \,.
\label{V_sym2}
\end{equation}
Also after the spontaneous symmetry breaking, a $\alpha v^2 R^2$ term appears in the action. This term will drive inflation and the scalar field variation $\xi$ will behave as a dark fluid.

\subsubsection{Inflation in Einstein frame}

Usually, as for example in the Higgs-inflation model, the symmetry breaking occurs after inflation but in our case it is the opposite. After symmetry breaking, which corresponds to $|\xi| \ll v$, the Universe can expand exponentially as in the $R^2$ model. Assuming that $\xi$ has a negligible effect during inflation the action \eqref{S2} becomes:
\begin{equation}
\mathcal{S}=\int d^4x \sqrt{-g}\left[\frac{1}{2\kappa^2}\left(R+\frac{\alpha v^2}{M_P^2} R^2 \right) \right].
\end{equation}        
This action corresponds to the $f(R) = R + R^2/(6M^2)$ theory with $6 M^2 = M_P^2/(\alpha v^2)$. Starobinsky inflation model is known to be a viable inflation scenario. Inflation occurs when $R \gg M^2$ and $H^2 \gg |\dot{H}|$, and the typical value of $M$ is $3\times 10^{-6} m_P$ with $m_P=\sqrt{8\pi} M_P$ \cite{DeFelice:2010aj}.

We make a conformal transformation to go to the Einstein frame, with
\begin{equation}
\Omega^2 = F(R) \,,
\end{equation}
where $F(R)$ is the derivative of $f(R)$. By defining an effective field $\psi$ as:
\begin{equation}
\frac{\psi}{M_P} =  \sqrt{\frac{3}{2}} \log F\,,
\end{equation}
the action becomes:
\begin{equation}
\mathcal{S} =\int d^4x \sqrt{-\tilde{g}}\left[\frac{1}{2\kappa^2}\tilde{R} -\frac{1}{2}\tilde{g}^{\mu \nu}\partial_{\mu}\psi\partial_{\nu}\psi- U(\psi) \right]\,,
\end{equation}
with
\begin{equation}
U(\psi) = \frac{M_P^2}{2} \frac{FR-f}{F^2} = \frac{M_P^4}{8\alpha v^2} \left[1 - \exp \left(-\sqrt{\frac{2}{3}} \frac{\psi}{M_P} \right) \right]^2\,.
\label{pot_inf}
\end{equation}
\begin{figure}[!t]
\centering
\includegraphics[width=15cm]{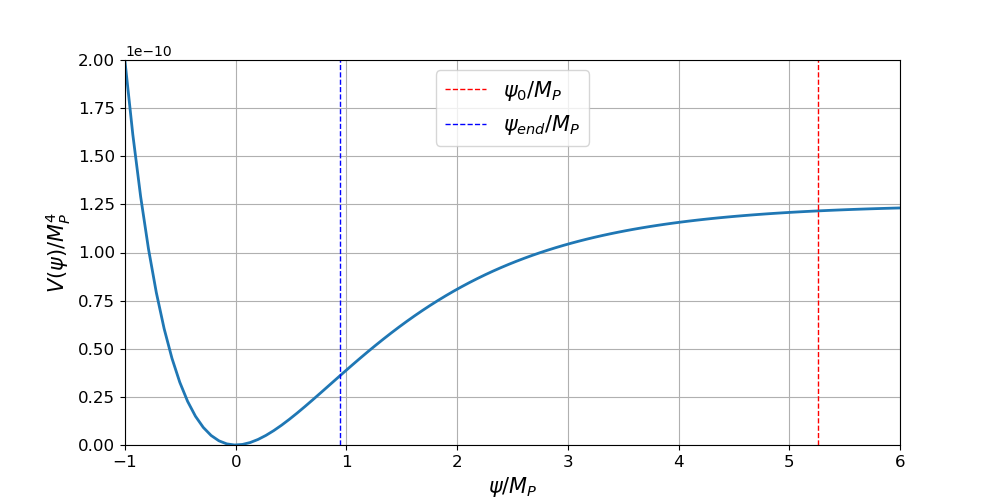}
\caption{Potential $U(\psi)$ defined in Eq.~\eqref{pot_inf} as a function of $\psi$. Inflation starts at the dashed red line and ends at the dashed blue line.\label{fig:pot}}
\end{figure}%
The tilde in the action denotes quantities in the Einstein frame. Chaotic inflation occurs when the potential dominates the evolution and changes slowly. 

\begin{figure}[!th]
\hspace*{-1.cm}$\begin{array}{cc}
 \includegraphics[width=8cm]{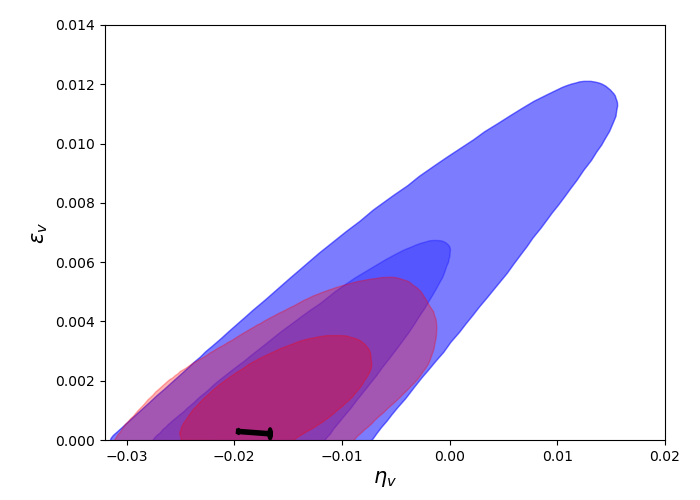} &
 \includegraphics[width=8cm]{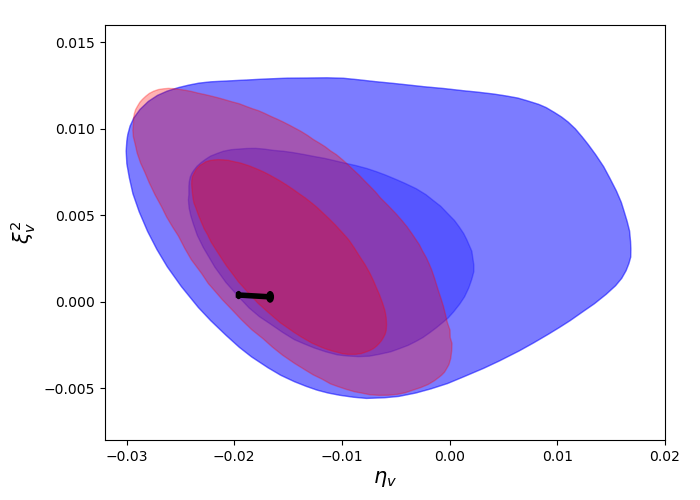} \\
 \includegraphics[width=8cm]{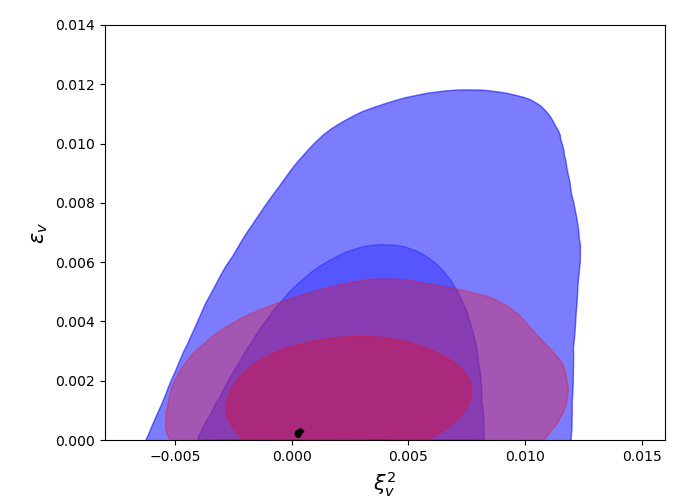} & 
 \includegraphics[width=8cm]{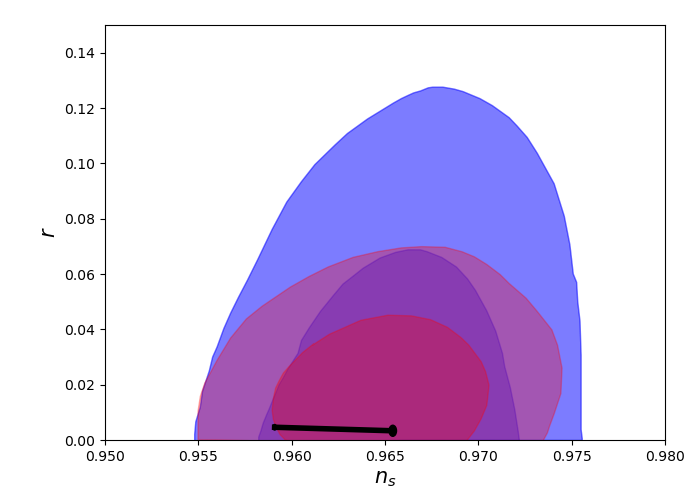}
\end{array}$
\caption{Marginalized joint two-dimensional $68\%$ and $95\%$ C.L. regions for the slow-roll parameters $(\epsilon_v,\eta_v,\xi^2_v)$, and the tensor to scalar ratio $r$ as a function of the scalar spectral index $n_s$. The two-dimensional constraints are obtained by the Planck collaboration \cite{Akrami:2018odb} using TT,TE,EE+lowE+lensing data (blue contours) and TT,TE,EE+lowE+lensing+BK15 data (red contours). The black lines correspond to the values of our inflation model for $50<N<60$. The big black dot is calculated with $N=60$ and the small one with $N=50$. \label{fig:slow_roll}}     
\end{figure}

The shape of the potential is shown in Fig.~\ref{fig:pot}, where one can see that the potential is relatively constant in the region $\psi \gg M_P$. The slow-roll parameters of this potential, which have to be small during inflation, are given by:
\begin{equation}
\begin{aligned}
\epsilon_v &= \frac{M_P^2}{2}\left(\frac{U'}{U}\right)^2 = \frac{4}{3} \left[ \exp \left(\frac{2\psi}{\sqrt{6}M_P}\right)-1 \right]^{-2} \displaystyle \,,\\
\eta_v &= M_P^2 \frac{U''}{U} = -\frac{4}{3} \frac{\exp \left(\frac{2\psi}{\sqrt{6}M_P}\right) - 2 }{\left[ \exp \left(\frac{2\psi}{\sqrt{6}M_P}\right)-1 \right]^{2} } \displaystyle \,,\\
\zeta^2_v &= M_P^4 \frac{U''' U'}{U^2} = \frac{16}{9} \frac{\exp \left(\frac{2\psi}{\sqrt{6}M_P}\right) - 4 }{\left[ \exp \left(\frac{2\psi}{\sqrt{6}M_P}\right)-1 \right]^{3} } \displaystyle \,.
\end{aligned}
\label{slow-roll}
\end{equation}
The end of inflation is characterized by $\epsilon_v=1$ so that
\begin{equation}
 \frac{\psi_{end}}{M_P} = \sqrt{\frac{3}{2}} \ln \left( 1 + \frac{2}{\sqrt{3}} \right) \simeq 0.94 \,.
\end{equation}
The number of e-folds is given by:
\begin{equation}
N = \frac{1}{M_P^2}\int_{\psi_{end}}^{\psi_{beg}}\frac{U}{U'}d\psi  = \frac{3}{4} \left[ \left( \exp \left( \sqrt{\frac{2}{3}}\frac{\psi_{beg}}{M_P}\right) - \sqrt{\frac{2}{3}}\frac{\psi_{beg}}{M_P} \right) - \left( \exp \left( \sqrt{\frac{2}{3}}\frac{\psi_{end}}{M_P}\right) - \sqrt{\frac{2}{3}}\frac{\psi_{end}}{M_P} \right) \right] \,,
\end{equation}
and one obtains for $N = 55$
\begin{equation}
\frac{\psi_{beg}}{M_P} \simeq \sqrt{\frac{3}{2}} \ln \left( \frac{4}{3}N \right) \simeq 5.26\,.
\end{equation} 
The scalar spectral index \cite{Baumann:2009ds} and the tensor to scalar ratio can also be obtained as $n_s=1-6 \epsilon_v + 2\eta_v$ and $r=16\epsilon_v$, respectively. The CMB observations by Planck set constraints on slow-roll parameters \cite{Ade:2015lrj}, which are presented in Fig.~\ref{fig:slow_roll}. Therefore our predictions are in agreement at the one sigma level with Planck observational data.

In order to obtain the value of $\alpha v^2$, we will use the amplitude of the power spectrum which is connected to the potential by  \cite{Lyth:1998xn}:
\begin{equation}
\delta_H^2 = \frac{4}{25} \mathcal{P}_{\mathcal{R}} = \frac{1}{150 \pi^2 M_P^4} \frac{V}{\epsilon_v}\,.
\end{equation}
The evaluation of this expression at the end of inflation, which corresponds to $N\simeq 55$ and $\delta_H \simeq 2\times 10^{-5}$, gives:
\begin{equation}
\alpha v^2 \simeq \frac{N^2}{144 \pi^2 \mathcal{P}^*_{\mathcal{R}}} \simeq 10^9\,.
\end{equation}
We have therefore a value for $\alpha v^2$, but $\alpha$ and $v$ are not constrained independently. We however made the assumption that the symmetry breaking occurs before inflation. This assumption imposes only $\xi < v$ and it is for this reason that we neglected all the terms in front of $R^2$ in Eq. \eqref{S2} except the constant term.
 
As mentioned at the beginning of the section, $\alpha v^2$ is related to the parameter $M$ of the $R^2$-inflation via $M=1/\sqrt{48\pi \alpha v^2}m_P \approx 3\times 10^{-6} m_P$. In the following, we will use $M$ instead of $\alpha v^2$.

\subsubsection{Reheating}

In this section, we study the production of particles in the Jordan frame, which can be created via the Unruh effect \cite{Mukhanov:2007zz} for which an observer in an accelerated frame can observe the emission of particles off the vacuum. Let us consider the usual reheating scenario after $R^2$ inflation \cite{DeFelice:2010aj,Akrami:2018odb}. As long as the created particles do not modify the evolution of the Universe, the action which describes the model is  still:
\begin{equation}
\mathcal{S} = \int d^4x \sqrt{-g} \left[ \frac{1}{2\kappa^2}\left( R + \frac{R^2}{6M^2} \right) \right]\,,
\end{equation}   
but the solutions \eqref{Sol_R2inf} are no longer valid. After inflation the Hubble parameter and scalar curvature are in a fast oscillating regime with $M(t-t_{os}) \gg 1$, where $t_{os}$ corresponds to the beginning of the oscillations. The solutions of the Einstein equations are then:
\begin{equation}
\begin{aligned}
& R \simeq - \frac{4M}{t-t_{os}} \sin \left( M(t-t_{os}) \right) \,, \\
& H \simeq \frac{4}{3(t-t_{os})}\cos^2 \left( \frac{M}{2}(t-t_{os}) \right) \,, \\
& \frac{a}{a_0} \simeq \left( t - t_{os}\right)^{2/3}\,.
\end{aligned}
\label{Sol_R2os}
\end{equation}
During this oscillating phase, radiation will be produced via the Unruh effect and the total radiation energy density is given by \cite{Mijic:1986iv,DeFelice:2010aj}:
\begin{equation}
\rho_r = \frac{g_*}{a^4} \int_{t_{os}}^t  \frac{Ma^4R^2}{1152\pi} dt\,,
\label{rad_density}
\end{equation}  
where $g_*$  is the number of relativistic degrees of freedom. At this period the radiation density encompasses (at least) all the Standard Model particles. The evolution of the radiation density is given by:
\begin{equation}
\frac{d\rho_r}{dt} + 4H\rho_r = \frac{g_*Ma^4R^2}{1152\pi}\,.
\end{equation}
The second term corresponds to a backreaction due to the expansion of the Universe. At a later stage, when $R$ becomes negligible and the right-hand side term in the previous equation vanishes, the total energy density evolves as $\rho_r \propto a^{-4}$. The integration of Eq.~\eqref{rad_density} is performed by considering the solution of Einstein equations given in~\eqref{Sol_R2os}. Without considering the backreaction of radiation, we obtain:
\begin{equation}
\rho_r \simeq \frac{g_*M^3}{240\pi}\frac{1}{t-t_{os}}\,.
\end{equation}
This density evolves slowly as compared to $H^2$. The radiation domination era begins at $t \simeq t_{os} + 10^3 M_P^2/(g_* M^3)$ and the reheating temperature can be approximated by \cite{DeFelice:2010aj,Akrami:2018odb}:
\begin{equation}
T_r \le 3\times 10^{16} g_*^{1/4} \left(\frac{M}{M_P} \right)^{3/2} \text{GeV}\,,
\end{equation}
where the $T_r$ is defined implicitly as the temperature at which $\rho_r = g_* \pi^2 T_r^4 /30$. 

\begin{figure}[!t]
\centering
\includegraphics[width=15cm]{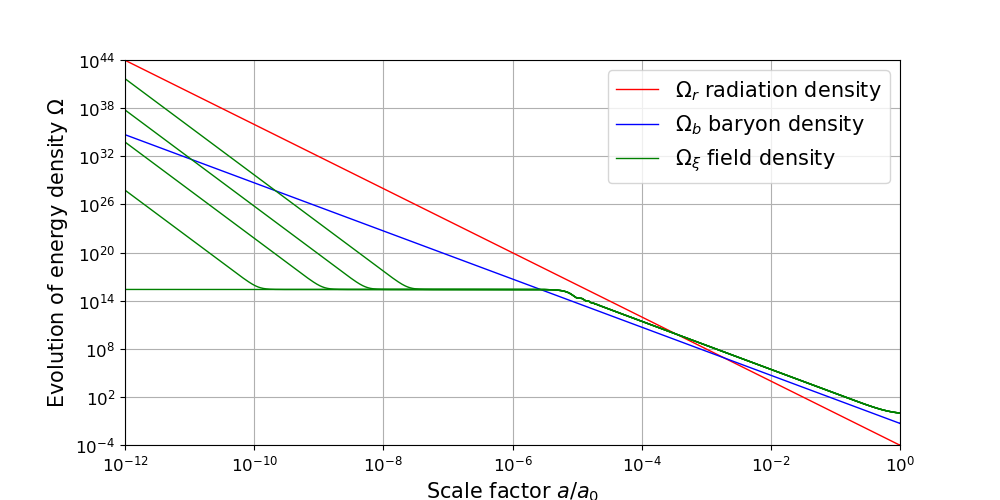}
\caption{Possible evolutions of the dark fluid density $\rho_{\xi}$ (green) depending on its initial density. The evolutions of the baryon density (blue) and radiation density (red) are the same as in the $\Lambda$CDM model.\label{fig:reheating}}
\end{figure}

As mentioned in the previous section the field $\xi$ exists before inflation but its density is suppressed by the expansion of the Universe and is negligibly small at the end of inflation. The reheating mechanism of the field $\xi$ is similar to the case of radiation. The energy density of $\xi$ is given by: 
\begin{equation}
\rho_{\xi} \simeq \frac{a_{os}^2}{a^6} \int_{t_{os}}^t  \frac{Ma^4R^2}{1152\pi} dt\,.
\end{equation} 
This density production is negligible compared to the radiation density since the latter is proportional to the large number of relativistic degrees of freedom. The evolution of the density of $\xi$ is shown in Fig.~\ref{fig:reheating} for different initial densities. One can notice that the scalar field density has no influence on the evolution of the Universe until it starts behaving like dark matter.

\subsection{Dark fluid behaviour}

After inflation the scalar curvature is small and the $R^2$ term can be safely neglected. The action \eqref{S2} therefore becomes:
\begin{equation}
\mathcal{S}=\int d^4x \sqrt{-g}\left[\frac{1}{2\kappa^2}R -\frac{1}{2}g^{\mu \nu}\partial_{\mu}\xi\partial_{\nu}\xi- V(\xi) \right] \,,
\end{equation}
with the potential given in Eq.~\eqref{V_sym2}.

As we have seen in the previous section the scalar field $\xi$ is expected to behave like a dark fluid. The scalar field $\xi$ can replace dark energy if:
\begin{equation}
V_0 = \frac{\Lambda}{\kappa} = 2.5\times 10^{-11} \text{ eV}^4\,,
\end{equation}
and dark matter if:
\begin{equation}
m \sim 10^{-22}\text{ eV}\,.
\end{equation}
In addition to the constant term and the mass term the potential contains a $\pm m^2/(2v) \xi^3$ term and a quartic term $m^2/(8v^2) \xi^4$. The effect of the quartic term has already been discussed in Section~\ref{sec:quartic}, and we have seen that $\lambda= m^2/2v^2$ has to be smaller than $10^{-98}$, which implies:
\begin{equation}
v > 7 \times 10^{26} \text{ eV}\,.
\end{equation}   
With such a large value of $v$, which is one order of magnitude below the Planck energy, the symmetry breaking will occur right after Planck time and well before inflation, and the term in $R^2$ can be safely neglected as we had assumed.

\begin{figure}[!t]
\centering
\includegraphics[width=15cm]{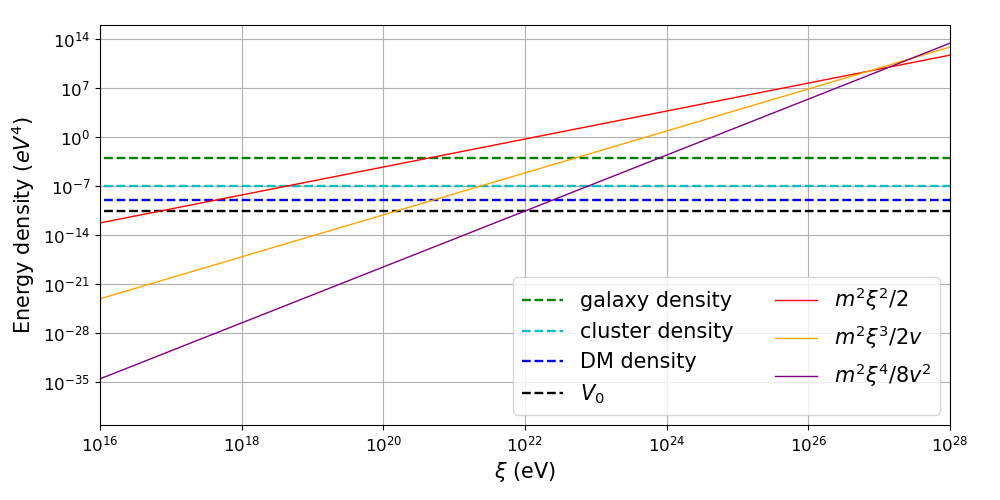}
\caption{Contributions to the scalar field energy density of the terms $m^2 \xi^2/2$ (red), $m^2 \xi^3/2v$ (yellow) and $m^2 \xi^4/8v^2$ (violet) in the potential~\eqref{V_sym2}, as a function of the value of the scalar field $\xi$. The horizontal line indicate the dark matter energy densities at different scales and the potential constant term $V_0$ corresponding to the cosmological constant.\label{fig:pot2}}
\end{figure}

Let us study the impact of the extra $\xi^3$ and $\xi^4$ terms of the potential. The scalar field $\xi$ can emulate dark matter if it oscillates quickly around its minimum. In such a case, neglecting the extra terms, the energy density of the dark fluid is:
\begin{equation}
\rho_{\xi} \simeq \frac{\dot{\xi}^2}{2} + \frac{m^2}{2}\xi^2 + V_0 = m^2 \langle \xi^2 \rangle  + V_0\,,
\end{equation}
where $m$ is fixed to $10^{-22}$ eV and $v = 7\times 10^{26}$ eV. Figure~\ref{fig:pot2} shows the contributions to the scalar field energy density from the different terms of the potential, as a function of the scalar field value $\xi$. The average dark matter energy densities in the recent Universe, in galaxies and in galaxy clusters are also shown for comparison. As can be seen in the figure, the cubic and quartic terms have negligible contributions as compared to the mass term of the potential. For example the average dark matter density in galaxies can be obtained for $\xi\sim 3\times 10^{20}$ eV. With this value we have: 
\begin{equation}
\begin{aligned}
\frac{m^2\xi^3}{2v}\left/\frac{m^2\xi^2}{2}\right. \simeq 5 \times 10^{-7}\,,\\
\frac{m^2\xi^4}{8v^2}\left/\frac{m^2\xi^2}{2}\right. \simeq 5\times 10^{-14}\,.
\end{aligned}
\end{equation}
The contributions of the cubic and quartic terms are even smaller at larger scales. Therefore, they do not affect the dark matter behaviour of the scalar field in galaxies and clusters, and the dark fluid behaves as fuzzy dark matter.

%%%%%%%%%%%%%%%%%%%%%%%%%%%%%%%%%%%

\section{Conclusions}

In this paper we have studied a scenario in which a scalar field does not only generate inflation, but also replaces dark matter and dark energy. We have shown that such a scalar field can exist provided a symmetry breaking occurs before inflation, and that its mass term is as tiny as $m \sim 10^{-22}$ eV.

In a generic way, we showed how we can extend the potential of dark fluid models. In dark fluid scenarios the scalar field oscillates quickly around its non-zero minimum during the matter-domination era, and the potential is dominated by a quadratic mass term. The shape of the potential on the other hand is unknown during the radiation-domination era and remains rather unconstrained. We showed that such a model can be consistent with the observations of the cosmological microwave background, the Lyman-$\alpha$ forest, galaxy rotational curves, and replaces both dark energy and dark matter.

We have also studied a triple unification model, relying on a single scalar field with a non-minimal gravitational coupling to the squared scalar curvature. The potential of this scalar field has been chosen to be a one-dimensional Mexican hat with two displaced minima plus a constant term. The constant term may also be replaced by a quintessence potential. After a discrete symmetry breaking similar to the one of the Higgs mechanism, a constant term appears in front of the squared scalar curvature, which can generate a $R^2$-inflation. At the end of inflation, the Standard Model particles will be produced via a reheating mechanism and the massive scalar field resulting from the symmetry breaking will also be reheated. This scalar field has the dark fluid properties and can simultaneously replace dark matter and dark energy. This scenario, in addition to replacing inflation, dark matter and dark energy with a single scalar field, leads to a dark matter behaviour similar to the one of fuzzy dark matter, which alleviates the cuspy halo and missing satellite problems. 

\bibliographystyle{h-physrev5}
\bibliography{biblio}

\begin{thebibliography}{10}

\bibitem{Aad:2012tfa}
ATLAS, G.~Aad {\em et~al.},
\newblock Phys. Lett. B {\bf 716}, 1 (2012), arXiv:1207.7214.

\bibitem{Chatrchyan:2012ufa}
CMS, S.~Chatrchyan {\em et~al.},
\newblock Phys. Lett. B {\bf 716}, 30 (2012), arXiv:1207.7235.

\bibitem{Zlatev:1998tr}
I.~Zlatev, L.-M. Wang, and P.~J. Steinhardt,
\newblock Phys. Rev. Lett. {\bf 82}, 896 (1999), arXiv:astro-ph/9807002.

\bibitem{Amendola:1999er}
L.~Amendola,
\newblock Phys. Rev. D {\bf 62}, 043511 (2000), arXiv:astro-ph/9908023.

\bibitem{Irsic:2017yje}
V.~Iršič, M.~Viel, M.~G. Haehnelt, J.~S. Bolton, and G.~D. Becker,
\newblock Phys. Rev. Lett. {\bf 119}, 031302 (2017), arXiv:1703.04683.
%%CITATION = ARXIV:1703.04683;%%

\bibitem{Hu:2000ke}
W.~Hu, R.~Barkana, and A.~Gruzinov,
\newblock Phys. Rev. Lett. {\bf 85}, 1158 (2000), arXiv:astro-ph/0003365.

\bibitem{Boyle:2001du}
L.~A. Boyle, R.~R. Caldwell, and M.~Kamionkowski,
\newblock Phys. Lett. B {\bf 545}, 17 (2002), arXiv:astro-ph/0105318.

\bibitem{Arbey:2003sj}
A.~Arbey, J.~Lesgourgues, and P.~Salati,
\newblock Phys. Rev. {\bf D68}, 023511 (2003), arXiv:astro-ph/0301533.
%%CITATION = ASTRO-PH/0301533;%%

\bibitem{Guth:1980zm}
A.~H. Guth,
\newblock Adv. Ser. Astrophys. Cosmol. {\bf 3}, 139 (1987).

\bibitem{Bilic:2001cg}
N.~Bilic, G.~B. Tupper, and R.~D. Viollier,
\newblock Phys. Lett. {\bf B535}, 17 (2002), arXiv:astro-ph/0111325.
%%CITATION = ASTRO-PH/0111325;%%

\bibitem{Arbey:2005fn}
A.~Arbey,
\newblock (2005), arXiv:astro-ph/0506732.
%%CITATION = ASTRO-PH/0506732;%%

\bibitem{Arbey:2006it}
A.~Arbey,
\newblock Phys. Rev. {\bf D74}, 043516 (2006), arXiv:astro-ph/0601274.
%%CITATION = ASTRO-PH/0601274;%%

\bibitem{Arbey:2008gw}
A.~Arbey,
\newblock Open Astron. J. {\bf 1}, 27 (2008), arXiv:0812.3122.

\bibitem{Linde:2002ws}
A.~D. Linde,
\newblock arXiv:hep-th/0205259.

\bibitem{Bastero-Gil:2015lga}
M.~Bastero-Gil, R.~Cerezo, and J.~G. Rosa,
\newblock Phys. Rev. D {\bf 93}, 103531 (2016), arXiv:1501.05539.

\bibitem{Liddle:2008bm}
A.~R. Liddle, C.~Pahud, and L.~Urena-Lopez,
\newblock Phys.\ Rev.\ D {\bf 77}, 121301 (2008), arXiv:0804.0869.

\bibitem{Liddle:2006qz}
A.~R. Liddle and L.~A. Urena-Lopez,
\newblock Phys. Rev. Lett. {\bf 97}, 161301 (2006), arXiv:astro-ph/0605205.

\bibitem{DeSantiago:2011qb}
J.~De-Santiago and J.~L. Cervantes-Cota,
\newblock Phys. Rev. D {\bf 83}, 063502 (2011), arXiv:1102.1777.

\bibitem{Starobinsky:1980te}
A.~A. Starobinsky,
\newblock Adv.\ Ser.\ Astrophys.\ Cosmol. {\bf 3}, 130 (1987).

\bibitem{Peebles:2000yy}
P.~J.~E. Peebles,
\newblock Astrophys. J. {\bf 534}, L127 (2000), arXiv:astro-ph/0002495.
%%CITATION = ASTRO-PH/0002495;%%

\bibitem{Arbey:2001qi}
A.~Arbey, J.~Lesgourgues, and P.~Salati,
\newblock Phys. Rev. {\bf D64}, 123528 (2001), arXiv:astro-ph/0105564.
%%CITATION = ASTRO-PH/0105564;%%

\bibitem{Hui:2016ltb}
L.~Hui, J.~P. Ostriker, S.~Tremaine, and E.~Witten,
\newblock Phys. Rev. D {\bf 95}, 043541 (2017), arXiv:1610.08297.

\bibitem{Bernal:2006ci}
A.~Bernal and F.~Siddhartha~Guzman,
\newblock Phys. Rev. D {\bf 74}, 103002 (2006), arXiv:astro-ph/0610682.

\bibitem{Clowe:2006eq}
D.~Clowe {\em et~al.},
\newblock Astrophys. J. Lett. {\bf 648}, L109 (2006), arXiv:astro-ph/0608407.

\bibitem{Arbey:2019cpf}
A.~Arbey and J.~F. Coupechoux,
\newblock JCAP {\bf 1911}, 038 (2019), arXiv:1907.04367.
%%CITATION = ARXIV:1907.04367;%%

\bibitem{Cembranos:2018ulm}
J.~A.~R. Cembranos, A.~L. Maroto, S.~J. Núñez~Jareño, and
  H.~Villarrubia-Rojo,
\newblock JHEP {\bf 08}, 073 (2018), arXiv:1805.08112.
%%CITATION = ARXIV:1805.08112;%%

\bibitem{Aghanim:2018eyx}
Planck, N.~Aghanim {\em et~al.},
\newblock (2018), arXiv:1807.06209.
%%CITATION = ARXIV:1807.06209;%%

\bibitem{Linder:2005ne}
E.~V. Linder and D.~Huterer,
\newblock Phys. Rev. {\bf D72}, 043509 (2005), arXiv:astro-ph/0505330.
%%CITATION = ASTRO-PH/0505330;%%

\bibitem{Tsujikawa:2013fta}
S.~Tsujikawa,
\newblock Class. Quant. Grav. {\bf 30}, 214003 (2013), arXiv:1304.1961.

\bibitem{Caldwell:2005tm}
R.~Caldwell and E.~V. Linder,
\newblock Phys. Rev. Lett. {\bf 95}, 141301 (2005), arXiv:astro-ph/0505494.

\bibitem{Arbey:2007vu}
A.~Arbey and F.~Mahmoudi,
\newblock Phys. Rev. {\bf D75}, 063513 (2007), arXiv:hep-th/0703053.
%%CITATION = HEP-TH/0703053;%%

\bibitem{Linder:2006uf}
E.~V. Linder,
\newblock J. Phys. A {\bf 40}, 6697 (2007), arXiv:astro-ph/0610173.

\bibitem{Frieman:1995pm}
J.~A. Frieman, C.~T. Hill, A.~Stebbins, and I.~Waga,
\newblock Phys. Rev. Lett. {\bf 75}, 2077 (1995), arXiv:astro-ph/9505060.

\bibitem{Akrami:2018odb}
Planck, Y.~Akrami {\em et~al.},
\newblock (2018), arXiv:1807.06211.

\bibitem{Linde:1983gd}
A.~D. Linde,
\newblock Phys. Lett. B {\bf 129}, 177 (1983).

\bibitem{Bezrukov:2007ep}
F.~L. Bezrukov and M.~Shaposhnikov,
\newblock Phys. Lett. {\bf B659}, 703 (2008), arXiv:0710.3755.
%%CITATION = ARXIV:0710.3755;%%

\bibitem{Kaiser:1994vs}
D.~I. Kaiser,
\newblock Phys. Rev. D {\bf 52}, 4295 (1995), arXiv:astro-ph/9408044.

\bibitem{Lyth:1998xn}
D.~H. Lyth and A.~Riotto,
\newblock Phys. Rept. {\bf 314}, 1 (1999), arXiv:hep-ph/9807278.
%%CITATION = HEP-PH/9807278;%%

\bibitem{Sotiriou:2008rp}
T.~P. Sotiriou and V.~Faraoni,
\newblock Rev. Mod. Phys. {\bf 82}, 451 (2010), arXiv:0805.1726.

\bibitem{DeFelice:2010aj}
A.~De~Felice and S.~Tsujikawa,
\newblock Living Rev. Rel. {\bf 13}, 3 (2010), arXiv:1002.4928.
%%CITATION = ARXIV:1002.4928;%%

\bibitem{Mijic:1986iv}
M.~B. Mijic, M.~S. Morris, and W.-M. Suen,
\newblock Phys. Rev. D {\bf 34}, 2934 (1986).

\bibitem{Baumann:2009ds}
D.~Baumann,
\newblock arXiv:0907.5424.

\bibitem{Ade:2015lrj}
Planck, P.~Ade {\em et~al.},
\newblock Astron.\ Astrophys. {\bf 594}, A20 (2016), arXiv:1502.02114.

\bibitem{Mukhanov:2007zz}
V.~Mukhanov and S.~Winitzki,
\newblock {\em {Introduction to quantum effects in gravity}} (Cambridge
  University Press, 2007).

\end{thebibliography}

\end{document}